\documentclass{article}
\usepackage{spconf,amsmath,graphicx}
\usepackage[misc]{ifsym}
\usepackage[cmyk]{xcolor}
\usepackage{multirow}
\usepackage{float}
\usepackage{booktabs}
\usepackage{cite}
\usepackage{hyperref}
\usepackage{verbatim}
\usepackage{algorithm}
\usepackage{algorithmic}
\usepackage{dcolumn}
\hypersetup{
  colorlinks=true,    
  linkcolor=black,     
  citecolor=black,    
  urlcolor=black 
}

\title{AudioTime: A Temporally-aligned Audio-text Benchmark Dataset}
\name{Zeyu Xie\textsuperscript{1,2},
        Xuenan Xu\textsuperscript{1}, 
        Zhizheng Wu\textsuperscript{2,3},
        \Letter\! Mengyue Wu\textsuperscript{1} \thanks{\Letter\! Corresponding author: Mengyue Wu, mengyuewu@sjtu.edu.cn }}
\address{
$^{1}$X-LANCE Lab, Shanghai Jiao Tong University,
$^{2}$Shanghai AI Lab\\ $^{3}$Chinese University of Hong Kong, Shenzhen\\
}

\begin{document}
%
\maketitle
\begin{abstract}
Recent advancements in audio generation have enabled the creation of high-fidelity audio clips from free-form textual descriptions.
However, temporal relationships, a critical feature for audio content, are currently underrepresented in mainstream models, resulting in an imprecise temporal controllability.
Specifically, users cannot accurately control the timestamps of sound events using free-form text. 
We acknowledge that a significant factor is the absence of high-quality, temporally-aligned audio-text datasets, which are essential for training models with temporal control. 
The more temporally-aligned the annotations, the better the models can understand the precise relationship between audio outputs and temporal textual prompts.
Therefore, we present a strongly aligned audio-text dataset, AudioTime. 
It provides text annotations rich in temporal information such as timestamps, duration, frequency, and ordering, covering almost all aspects of temporal control. 
Additionally, we offer a comprehensive test set and evaluation metric to assess the temporal control performance of various models. 
Examples are available on the \href{https://zeyuxie29.github.io/AudioTime/}{\textcolor{cyan}{\textit{AudioTime-Demo}}}.
\end{abstract}
\section{Introduction}
\label{sec:intro}
Audio generation tasks have garnered significant attention recently, with two diverse research directions.
One focuses on enhancing the quality and fidelity of audio generated from free text. 
With advancements in self-supervised representation learning, deep generative frameworks, and multimodal pretraining tasks, several systems~\cite{kreuk2022audiogen, yang2023diffsound, liu2023audioldm1, liu2023audioldm2, ghosal2023text, majumder2024tango, huang2023make1, huang2023make2} have been proposed to generate high-quality audio segments, prioritizing the accuracy of events described in the text. 
The other direction focuses on the alignment of generated sounds, where temporal alignment is an important trait for audio as a sequential signal.
Previous works have extracted temporal pivots for audio generation through text preprocessing~\cite{huang2023make2}, integrating additional input~\cite{guo2023audio}, and extracting from video-audio alignment~\cite{comunita2024syncfusion}, etc.
\textbf{However, achieving fine-grained control using only free-text remains less investigated.}
Previous works either do not focus on temporal alignments~\cite{kreuk2022audiogen, yang2023diffsound, liu2023audioldm1, liu2023audioldm2, ghosal2023text, majumder2024tango, huang2023make1}, achieve only coarse-grained alignment~\cite{huang2023make2}, or require additional information~\cite{guo2023audio, comunita2024syncfusion}.
Achieving free-text time control is essential for understanding more information described, not just events; otherwise, misalignments such as omissions, confusion, and monotony may occur  (see Section~\ref{ssec:exp-analysis}).
Therefore, this work aims to improve temporal alignment by proposing a benchmark dataset and a performance metric for evaluation.

\begin{figure}[tbp]
  \centering
  \centerline{\includegraphics[width=1\linewidth]{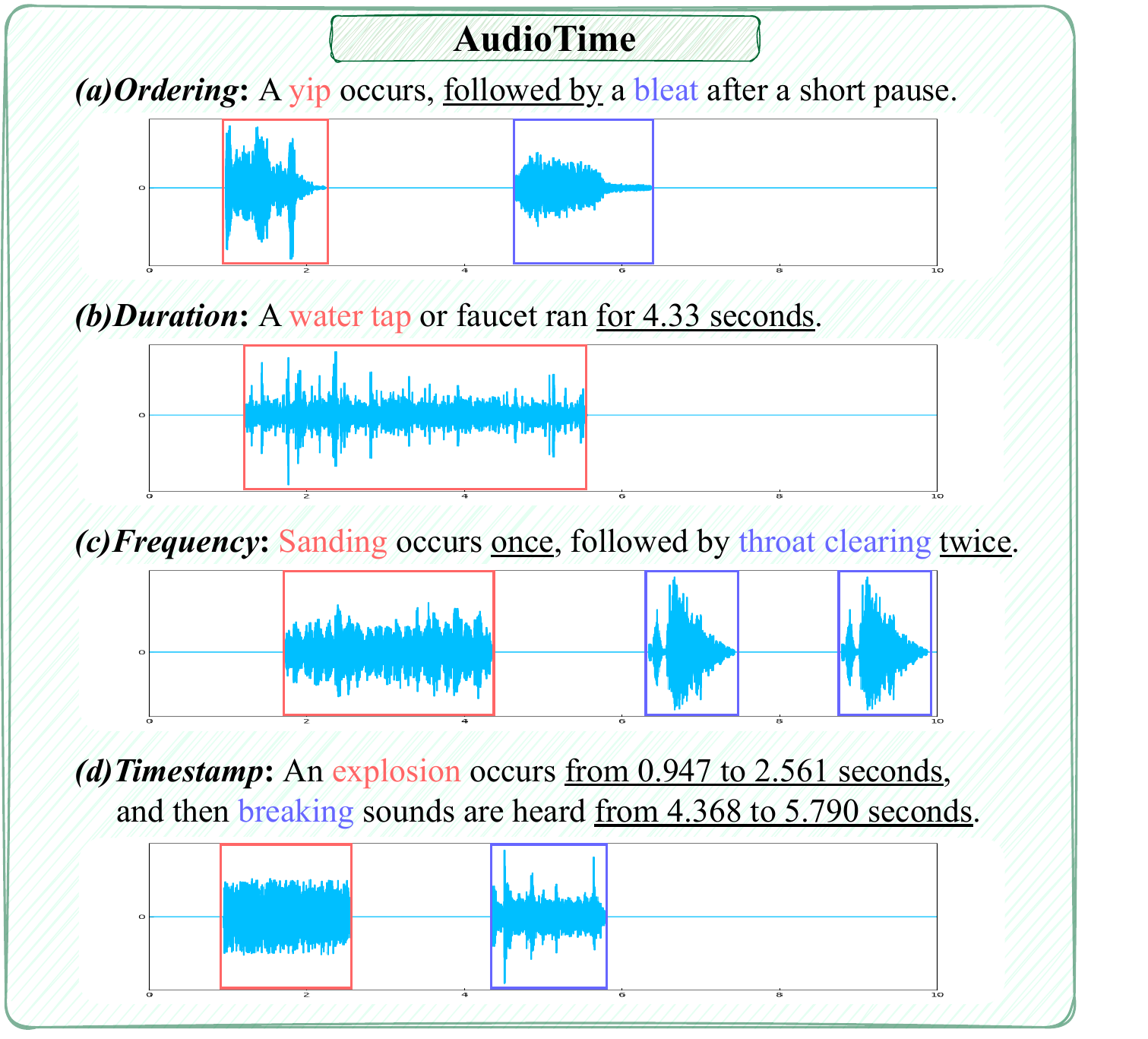}}
  \caption{Temporally-aligned audio-text samples in AudioTime.
  }
  \label{fig:intro_samples}
\end{figure}

\begin{figure*}[htbp]
  \centering
  \centerline{\includegraphics[width=1\textwidth]{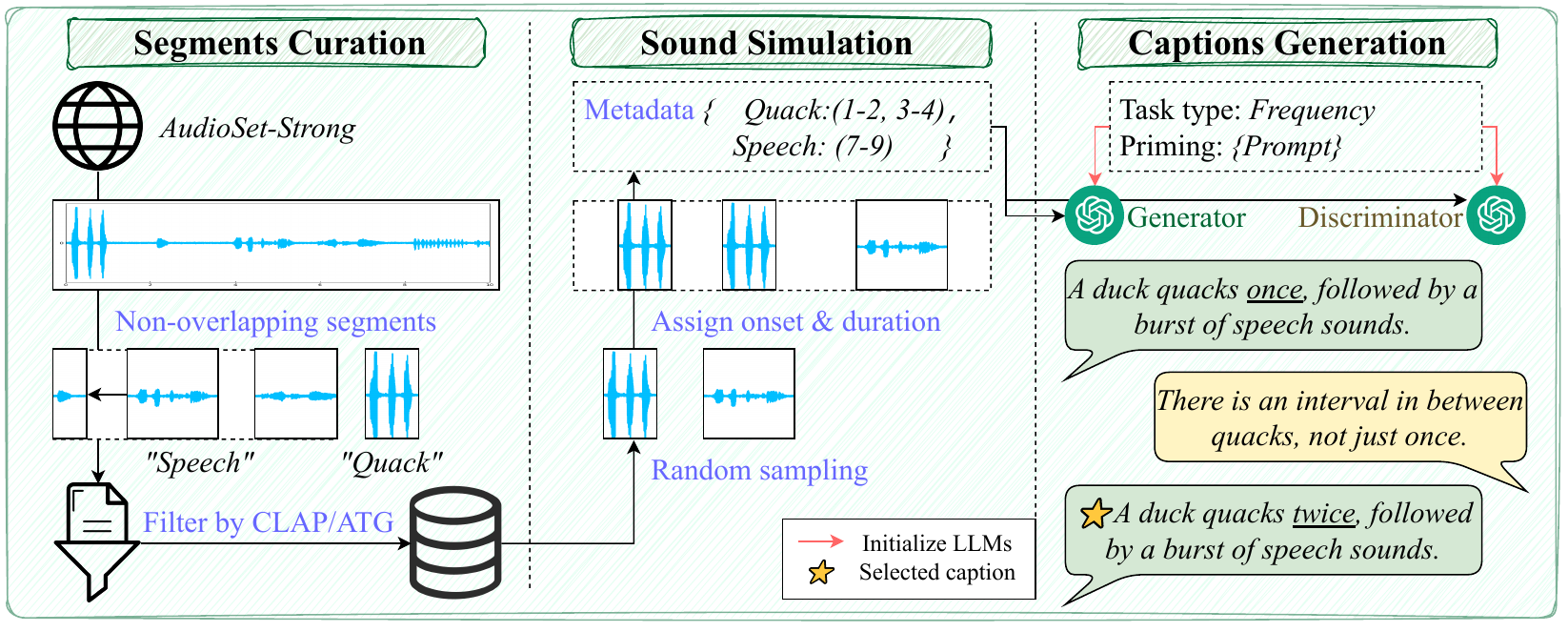}}
    %
    \caption{
    Construction pipeline: (1) Acquire clean segments from AudioSet-Strong and filtering using CLAP and ATG models; (2) Simulate audio clips with the Scaper tool and record metadata; (3) Generate captions using two agentic LLMs.
    }
    \label{fig:simulation_pipeline}
\end{figure*}

One of the reasons for the inability to achieve fine-grained control using only free-text is that the corpora are coarse-grained.
Naturally, the closer the annotations are temporally aligned, the more effectively models can comprehend the exact correspondence between audio and textual prompts.
%
By analyzing the largest available audio-text dataset, Audiocaps~\cite{kim2019audiocaps}, we observe that the captions usually provide only weakly supervised signals and lack precise annotations.
Specifically, the temporal relations in text and audio are loosely connected, such that specific durations, occurrence frequencies, and precise timestamps are difficult to ascertain (See Section~\ref{ssec:simulation}).
\textbf{Such coarse-grained annotations complicates the automated learning of temporal alignment.}

However, manually fine-grained temporal annotation is highly time-consuming and prone to substantial disagreement rates, given the inherent difficulty in precise time labeling.
Therefore, we design an automated pipeline to construct dataset \textbf{AudioTime}\footnote{The data is available at \href{https://github.com/zeyuxie29/AudioTime}{\textcolor{cyan}{\textit{AudioTime}}}.}, with detailed temporal attributes as supervised signals.
It eliminates the need for manual annotation, and facilitates cost-effective and time-efficient scalability. 
The pipeline involves three steps: segments curation, sound simulation, and captions generation, as shown in Figure~\ref{fig:simulation_pipeline}.
To facilitate comprehensive evaluation, we also propose \textbf{STEAM}\footnote{The evaluation script is available at \href{https://github.com/zeyuxie29/AudioTime/tree/main/STEAMtool}{\textcolor{cyan}{\textit{STEAMtool}}}.} (\textbf{S}trongly \textbf{TE}mporally-\textbf{A}ligned evaluation \textbf{M}etric), to assess the temporal control ability of models based on free text. 
Via STEAM, we analyze the performance of several influential models.
Our contributions encompass the following:  
\begin{enumerate}
\item Designing an automated pipeline to construct temporally aligned audio-text dataset AudioTime; 
\item Providing STEAM to evaluate the temporal control
ability of models;
\item Analyzing the temporal control performance of currently influential models.
\end{enumerate}

\section{AudioTime}
\label{sec:audiocapstrong}
To avoid the time-consuming nature and high disagreement rate of human annotation and to achieve scalability, a low-cost and efficient automated construction pipeline is introduced. 
It includes: (1) acquiring and filtering data segments; (2) simulating audio and organizing metadata; and (3) generating captions using agentic large language model (LLM).

\subsection{Audio Segments Curation}
\label{ssec:segment}
Non-overlapping segments are required to ensure annotation quality. 
However, target events in current datasets often overlap with other sounds (e.g., noise or other events)~\cite{xu2024detailed}, necessitating accurate localization and isolation.
We isolate \textbf{single-sound segments} from the AudioSet-Strong~\cite{gemmeke2017audio} dataset, as it provides on- \& off-set for events and the non-overlapping segments can be accurately located.
These segments contain clean single occurrence with precise duration annotations.


Since AudioSet is collected from the web, using events as queries to retrieve data may result in erroneous data, such as ``dog bark" actually being a person imitating a dog. 
These data need to be filtered out to ensure the quality of curated segments.
A contrastive language-audio pre-training (CLAP) model\footnote{\url{https://github.com/LAION-AI/CLAP}}~\cite{laionclap2023} is used to compute the similarity between sound event labels and one-occurrence segments, filtering out segments below predetermined threshold $0.3$. 
Subsequently, an audio-text grounding (ATG) model\footnote{\url{https://github.com/wsntxxn/TextToAudioGrounding}}~\cite{xu2024towards} uses the event label as a query to filter out segments below threshold $0.6$. 
The curation statistics are summarized in Table~\ref{tab:curation}.

\begin{table}[htbp]
\renewcommand{\arraystretch}{1}
    \centering
    \small
    \caption{ 
    Curation statistics. ``C" and ``S" denote categories and segments, respectively.
    The percentage (\%) represents the proportion remaining after filtering.
    }
    \begin{tabular}{c|cc|cc}
    \toprule
      \multirow{2}{*}{}&\multicolumn{2}{c|}{Extracted }&\multicolumn{2}{c}{Filtered}  \\
      &\# C&\# S &\# C&\# S \\
    \midrule
    Training   & 309  & 7,098 & 195 (63.1\%) & 3,392 (47.8\%) \\
    Evaluation  & 244 & 2,671 &143 (58.6\%)   &1,206 (45.2\%)\\
    \bottomrule
    \end{tabular}
    \label{tab:curation}
\end{table}

\begin{table*}[t]
\renewcommand{\arraystretch}{1}
    \centering
    \small
    \caption{ 
    Data statistics. 
    The asterisk ($\ast$) indicates the number obtained from the AudioSet ontology and annotations, with subclass merged into parent class to eliminate duplicates. 
    The double asterisk ($\ast \ast$) denotes the number detected by grounding model.
    The numbers before and after the slash (/) represent the total count and the average per clip, respectively.
    }
    \begin{tabular}{c|cc|ccc|ccc}
    \toprule
    \multirow{2}{*}{Dataset}&\multirow{2}{*}{Signal}&\multirow{2}{*}{Metadata}&\multicolumn{3}{c|}{Training }&\multicolumn{3}{c}{Evaluation}  \\
    & & &\# clips & \# events & \# occurrences & \# clips & \# events & \# occurrences \\
    
    \midrule
    AudioCaps & - & - &49501& 83724 / 1.69$\ast$ &228363 / 4.61$\ast \ast$& 964 & 1628 / 1.69$\ast$ & 4313 / 4.47$\ast \ast$ \\
    
    \midrule
    \multirow{4}{4em}{AudioTime} & Ordering  &On- \& off-set& \multirow{4}{*}{5000}& 10000 / 2.0& 17406 / 3.48& \multirow{4}{*}{500} & 1000 / 2.00 &  1848 / 3.70\\
    
    & Duration  &Duration& & 8163 / 1.63& 8163 / 1.63&  &898 / 1.80 &  898 / 1.80\\
    
    & Frequency &Onset& & 7747 / 1.55& 11410 /
2.28 &  & 841 / 1.68 & 1266 / 2.53\\
    
    & Timestamp & On- \& off-set& &7685 / 1.54& 10714 / 2.14&  & 827 / 1.65 &  1183 / 2.37\\

    \bottomrule
    \end{tabular}
    \label{tab:metadata}
\end{table*}
\subsection{AudioTime Sound Simulation}
\label{ssec:simulation}
To explore the real-world temporal paradigm, we analyze the currently largest audio-text dataset, AudioCaps~\cite{kim2019audiocaps},
and summarize the embedded temporal relations.
4 different types of temporal
relations are manifested through human annotation, namely (a) ordering: \textit{``Hissing \underline{followed by} clanking dishes and speech"}; (b) duration: \textit{``A train blows its horn \underline{for a long time}"}; (c) frequency: \textit{``A vehicle approaches and blows a horn \underline{multiple times}"}; and (d) position: \textit{``Continuous applause with whistle blowing \underline{at the end}"}. 

We simulate these 4 scenarios following the real-world paradigm: (a) ordering: the audio contains two sequentially occurring distinct events, each event occurring any number of times; (b) duration: any number of events occur, each event happening once; (c) frequency and (d) timestamp: any events occur any number of times.
The Scaper provides a convenient tool for simulating audio based on single-sound segments~\cite{salamon2017scaper}.
We randomly select segments and specify their onset, duration, and other details. 
Temporal alignment is enhanced through \textbf{metadata} recording, as illustrated in Table~\ref{tab:metadata}.
\subsection{AudioTime Captions}
\label{ssec:caption}
In addition to audio clips with diverse temporal relations, a temporal-aligned dataset requires corresponding captions that reflect the detailed temporal information using precise descriptors.
The strong natural language processing capabilities of LLM can be transferred for caption generation.
We employ two GPT models \cite{achiam2023gpt}, one acting as a generator and the other as a discriminator.
To efficiently scale the dataset, the training set is generated using GPT-3.5, while the test set ensures greater accuracy by utilizing GPT-4.

Specifically, (1) the generator receives the task type and the metadata to generate a specified temporally aligned caption. 
(2) The discriminator receives an additional input, the output from the generator.
It is responsible for evaluating whether the caption generated by the generator meets the given requirements. 
(3) If the caption does not meet the criteria, the discriminator should provide feedback to generator. 
This iterative process continues until the discriminator deems the output satisfactory, as depicted in Algorithm~\ref{alg:iterative_captioning}.

\begin{algorithm}
\caption{Iterative Temporal Captioning Generation}
\label{alg:iterative_captioning}
\begin{algorithmic}[1]
\REQUIRE Task type $T_k$, corresponding metadata $M_i$
\ENSURE Temporally-aligned caption $C_i$ 

\STATE Initialize $C_i \leftarrow \emptyset$, $feedback \leftarrow \emptyset$
\STATE Context setting for the generator and discriminator
\REPEAT
    \STATE $C_i \leftarrow$ Generator($T_k$, $M_i$, $feedback$) 
    \STATE $feedback \leftarrow$ Discriminator($T_k$, $M_i$, $C_i$) 
\UNTIL{$feedback$ indicates success}
\RETURN $C_i$

\end{algorithmic}
\end{algorithm}

\subsection{AudioTime Data Statistics}
The AudioTime comprises comprehensive temporally-aligned audio-text data, as illustrated in Table~\ref{tab:metadata}. 
There are a total of 4 temporal signals, each consisting of 5000 training and 500 test instances. 
Each instance includes (1) an audio clip of up to 10 seconds in length, (2) precise metadata corresponding to the signal, and (3) a temporally precise annotated caption.

\section{Temporal Control Evaluation}
\label{sec:eval}
To better test and compare the temporal controllability of models, we propose STEAM (Strongly TEmporally-Aligned evaluation Metric).
STEAM is a text-based metric that  evaluates \textbf{whether the audio segments follow the temporal description specified by the AudioTime test captions.}
An audio-text grounding model~\cite{xu2024towards} is employed to detect the on- \& off-set of events in audio with threshold $0.5$. 
STEAM assesses control performance based on detected timestamps and the control signal provided in the text description.

(a) Ordering: To determine whether the audio generates events A and B in the specified order, quantified by the \textbf{error rate}.
The definition of the temporal relationship between events follows Xie et al.~\cite{xie2023enhance}.
Specifically, the onset of A should occur before B.
If there is overlap, it should not exceed half the duration of the shorter event (otherwise, they are considered simultaneous). 

(b) Duration / (c) frequency: Calculate the absolute error between the event duration/frequency in the generated audio and the value specified in the text, averaged over the total number of events, denoted as \textbf{L$_1^{\text{second}}$}/\textbf{L$_1^{\text{freq}}$}:
\begin{equation}
    \text{L$_1$} = \frac{1}{N*E}\sum_{n=1}^{N}{\sum_{e=1}^{E}{|\#specified - \#detected|}}
\end{equation}
where $N$ and $E$ denote the number of samples and events, respectively.

(d) Timestamp: To measure the accuracy of controlling audio timestamps.
\textbf{F1$_{\text{segment}}$}~\cite{mesaros2016metrics}, a common metric in sound event detection task, is calculated based on the detected and specified on- \& off-set.


\section{Experiment}
\label{sec:exp}
Using the AudioTime and the metric STEAM, we test some currently influential audio generation models\footnote{AudioLDM2: https://github.com/haoheliu/audioldm2;\\Amphion: https://github.com/open-mmlab/Amphion;\\Make-An-Audio2: https://github.com/bytedance/Make-An-Audio-2;\\Tango2:https://huggingface.co/declare-lab/tango2.}. 
All models are used with default settings, with seeds set to $0$ if not specified.


\subsection{Result}
\label{ssec:exp-result}
The ability to generate temporally-aligned audio based on free text is quantitatively compared, as shown in Table~\ref{tab:result}.
Overall, current temporal control generation based on free text is not sufficiently advanced. 
Current models primarily focus on generating accurate sound events without fine-grained temporal control,
so there is a certain gap from the reference.
The main reasons are: (1) The encoders in current models do not explicitly understand fine-grained temporal control within free text; (2) They are trained on corpora lacking strong temporally-aligned audio-text datasets, thus unable to automatically learn corresponding alignment relationships.

In comparison, Make-An-Audio2 employs LLM to perform coarse-grained deconstruction of free text descriptions, transforming them into a format such as ``$<A \& start>$ @ $<B \& end>$", enabling the model to roughly understand the occurrence position of events. 
As a result, it achieves the best performance in terms of ordering and frequency control.
Inspired by this, transferring knowledge from LLM to assist in audio control generation is an effective method.

\begin{table}[t]
\renewcommand{\arraystretch}{1}
    \centering
    \small
    \caption{ 
    Control performance. GT: AudioTime-Test; A: AudioLDM2; Am: Amphion; M: Make-An-Audio2; T: Tango2.
    }
    \begin{tabular}{c|cccc}
    \toprule
     Signal & Ordering & Duration & Frequency &Timestamp\\
    Metrics &  Error rate $\downarrow$& L$_1^{\text{second}}$ $\downarrow$  &L$_1^{\text{freq}}$ $\downarrow$&F1$_{\text{segment}}$ $\uparrow $ \\

    \midrule
     GT   & 0.220  & 0.792 & 0.347 & 0.910 \\
    \midrule
    A  &0.960  & 3.404 & 1.639 & 0.543\\
     
    Am     & 0.946 & \textbf{3.078} &1.471 &0.386 \\
     
    M  & \textbf{0.758} & 3.400  & \textbf{1.415} & 0.559 \\
    
    T  & 0.858 & 3.695 & 1.521 & \textbf{0.609} \\

    \bottomrule
    \end{tabular}
    \label{tab:result}
\end{table}

\subsection{Analysis}
\label{ssec:exp-analysis}
We further analyze the shortcomings of current models when dealing with rich textual content and explain the importance of parsing temporal control signals.

\textbf{Event omission} Current models often miss events when faced with input text containing multiple events, resulting in omitted events in the generated audio. 
For instance, when there is a requirement for event A followed by event B, the A event often fills the entire clip segment. 
This is typically due to a bias towards earlier occurring events and the insensitivity of event ordering.

\textbf{Event confusion} When the required sound events in the text share similar characteristics, current models tend to generate a confused intermediate-state event. 
For example, for events like ``male speech" and ``baby crying", the model may generate a confused event resembling ``baby speech". 

\textbf{Monotonous repetition} Another drawback of current models is that the generated audio content often lacks variation in loudness and rhythm, resulting in monotony. 
Especially when confronted with a single sound event, such as generating repetitive ``dog barks" that are similar dozens of times, this can lead to distortion in the audio.

One measure to alleviate the above issues is to achieve precise text-controlled generation. 
If the model can recognize strong alignment in the text regarding ordering, timestamps, etc., it can more easily generate different events correctly, thereby reducing omission and confusion.
Similarly, duration or frequency information can effectively control the length or occurrence frequency of events, ensuring its authenticity.
Therefore, we aim to promote the achievement of these goals by proposing the AutioTime dataset and STEAM.

\section{Conclusion}
\label{sec:conclusion}
Current audio generation models can now produce indistinguishable audio based on free text descriptions. 
There have also been efforts to control audio generation at a coarse level
However, there has been no related research on fine-grained control based on free text.
This work introduces the AudioTime dataset from the perspective of temporally controlled generation based on free text. 
It provides temporally-aligned text annotations, aiming to enhance the strong alignment between audio and text data. 
The dataset is constructed using a fully automated process, which involves segment curation, sound simulation, and caption generation. 
Additionally, to comprehensively measure the temporal control performance of generative models, we propose the evaluation metric STEAM. 
We conduct experiments on several influential models, identify some weaknesses, and analyze the importance of temporal control in audio generation.




\vfill\pagebreak

\bibliographystyle{IEEEbib}
\bibliography{strings,refs}

\end{document}